\documentclass{aa}
\usepackage{graphicx}
\usepackage{txfonts}
\usepackage{hyperref}

\newcommand*\diff{\mathop{}\!\mathrm{d}}
\bibpunct{(}{)}{;}{a}{}{,}

\begin{document}

   \title{Global properties of the  light curves of magnetic, chemically peculiar stars as a testbed for the existence of dipole-like symmetry in surface structures}
   \titlerunning{The global properties of magnetic chemically peculiar star light curves}
   \author{M. Jagelka
          \inst{1}
          \and
          Z. Mikul\'{a}\v{s}ek
          \inst{1}
          \and
          S. H\"ummerich
          \inst{2, 3}
          \and
          E. Paunzen
          \inst{1}
          }

   \institute{Department of Theoretical Physics and Astrophysics, Masaryk University, Brno, Kotl\'{a}\v{r}sk\'{a} 2, CZ-611 37 Brno, Czech Republic\\
              \email{jagelka@mail.muni.cz}
        \and
                American Association of Variable Star Observers (AAVSO), Cambridge, USA
        \and
                Bundesdeutsche Arbeitsgemeinschaft f\"ur Ver\"anderliche Sterne e.V. (BAV), Berlin, Germany
             }

   \date{Preprint online version: February 1, 2019}

  \abstract
   {Magnetic, chemically peculiar stars are known for exhibiting surface abundance inhomogeneities (chemical spots) that lead to photometric and spectroscopic variability with the rotation period. It is commonly assumed that the surface structures are causally connected with the global magnetic field that dominates the photospheric and subphotospheric layers of these stars. As a rule, the observed magnetic fields show a simple dipole-like geometry, with the magnetic axis being noncollinear to the rotational one.}
   {The present study aims at detecting underlying patterns in the distribution of photometric spots in a sample of 650 magnetic, chemically peculiar stars and examines their link to the magnetic field topology.}
   {Photometric time-series observations from the ASAS-3 archive were employed to inspect the light-curve morphology of our sample stars and divide them into representative classes described using a principal component analysis. Theoretical light curves were derived from numerous simulations assuming different spot parameters and following the symmetry of a simple dipole magnetic field. These were subsequently compared with the observed light curves.}
   {The results from our simulations are in contradiction with the observations and predict a much higher percentage of double-wave light curves than is actually observed. We thereby conclude that the distribution of the chemical spots does not follow the magnetic field topology, which indicates that the role of the magnetic field in the creation and maintenance of the surface structures may be more subsidiary than what is  predicted by theoretical studies.}
   {}


   \maketitle

\section{Introduction} \label{intro}
Chemically peculiar (CP) stars are members of the upper part of the main sequence (spectral types B to F). They are generally slow rotators and are characterized by spectral peculiarities that indicate unusual atmospheric abundances of some chemical elements. This is due to the effects of radiative diffusion and gravitational settling \citep{2009A&A...495..937L,2010A&A...516A..53A,2015MNRAS.454.3143A}, which result in some elements being lifted to the outer parts of the stellar atmosphere while others settle down into the lower parts.

Some groups of CP stars (the so-called Ap/CP2 stars and the He-weak/CP4 stars; cf. \citealt{1974ARA&A..12..257P}) exhibit global magnetic fields with strengths up to several tens of kilogauss \citep{2007A&A...475.1053A} and are characterized by surface abundance inhomogeneities (`chemical spots').\footnote{For convenience, these stars are referred to hereafter as magnetic, chemically peculiar (mCP) stars.} Flux is redistributed in these spots (e.g., \citealt{1970ApJ...161..685P,1973ApJ...179..527M,2013A&A...556A..18K}) and, as the star rotates, the changing viewing angle results in periodic spectroscopic and photometric variations. Stars exhibiting this kind of rotational variability are traditionally referred to as $\alpha^2$ Canum Venaticorum (ACV) variables.

While a precise understanding of the processes involved in the formation of the surface structures is still missing \citep{2013A&A...556A..18K}, it is believed that the magnetic field plays a decisive role \citep{2009A&A...495..937L,2010A&A...516A..53A,2015MNRAS.454.3143A}. The present study aims at detecting underlying patterns in the distribution of photometric spots and examines their link to the magnetic field topology.

\section{Analysis of observations}

\subsection{Data characteristics}

The original data for our research were gleaned from the third phase of the All Sky Automated Survey (ASAS) project, which focused on continuous photometric coverage of the southern sky and part of the northern sky with the goal of detecting any kind of photometric variability. During the third phase of the project (ASAS-3), observations were acquired in Johnson \textit{V} \citep{2005AcA....55..275P}. In total, 11\,509 variable stars brighter than $V=15$\,mag were cataloged by the ASAS team, of which 7\,310 remained unclassified. \citet{2015A&A...581A.138B} used ASAS-3 data to search for new photometric variables among known mCP stars from the catalogue of \citet{2009A&A...498..961R} and identified 316 stars with photometric characteristics of ACV variables. Using refined search criteria, \citet{2016AJ....152..104H} subsequently discovered another 334 ACV variables in these data.

As basis for the present investigation, we employed the original ASAS-3 data of the 650 ACV variables identified by \citet{2015A&A...581A.138B} and \citet{2016AJ....152..104H}. We also used their ephemerides, such as the period and Julian date of photometric maximum. ASAS-3 data contain information about Julian Date, brightness using different photometric apertures, corresponding errors, and quality of measurements. We used measurements from the aperture with the lowest errors for a specific star following the methodology of \citet{2005AcA....55..275P}. For the brightest stars ($V<9$\,mag), the largest aperture (6 pixels in diameter) was employed; for faint stars ($V>12$\,mag), we used the smallest aperture (2 pixels in diameter). Furthermore, lower-quality measurements  \citep[indicated by flags `C' and `D'; for details see][]{2005AcA....55..275P} were excluded from the analysis.

\subsection{Principal component analysis of the light curves}

According to the shape of the light curves of the mCP stars studied by \citet{2015A&A...581A.138B} and \citet{2016AJ....152..104H}, we can divide them into simple categories: single-wave, double-wave, and symmetric or asymmetric light curves. To confirm and mathematically describe these classes, we used a principal component analysis (PCA). This method transforms the original data into a system with new coordinates, which are chosen such that they follow the most common patterns in the data. This coordinate system is based on the combination of the eigenvectors with large eigenvalues. Therefore, we can create principal components and use them to describe the light curve data as their linear combinations. For a detailed description of the modifications of PCA, we refer to \citet{2007A&AT...26...63M}.

\begin{figure}
\begin{center}
\includegraphics[width=0.45\textwidth]{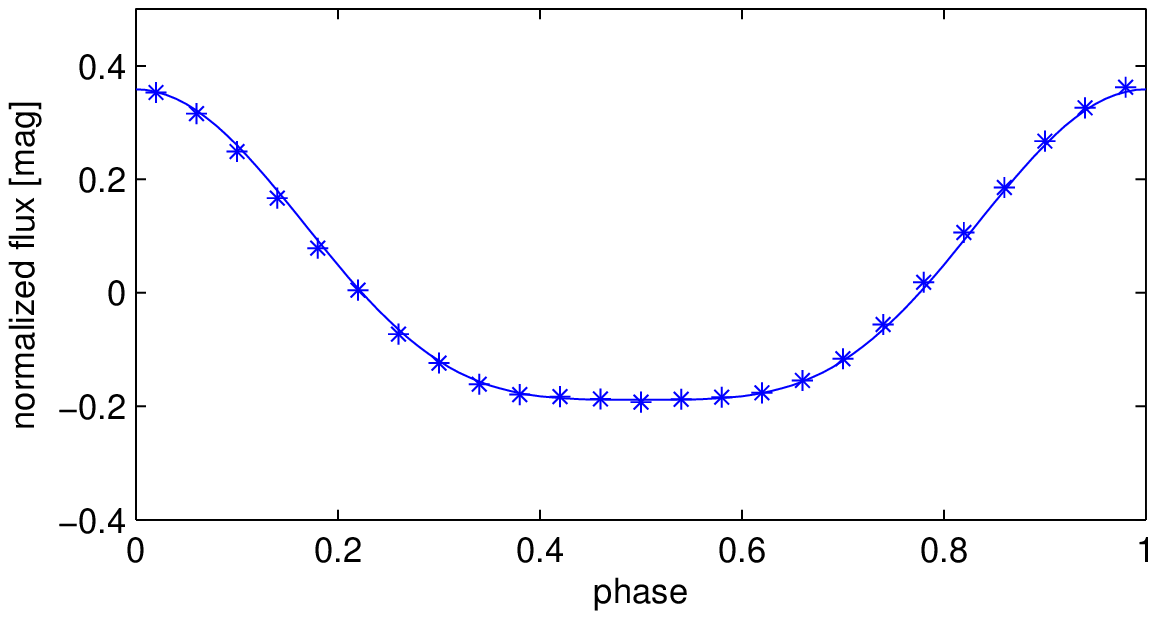}
\includegraphics[width=0.45\textwidth]{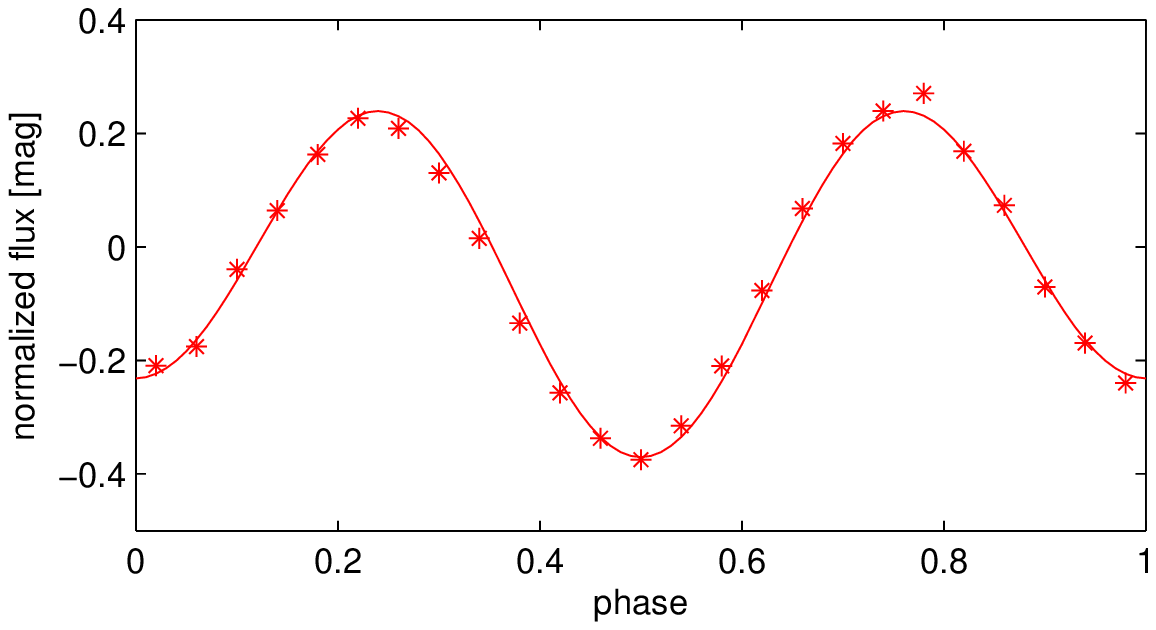}
\includegraphics[width=0.45\textwidth]{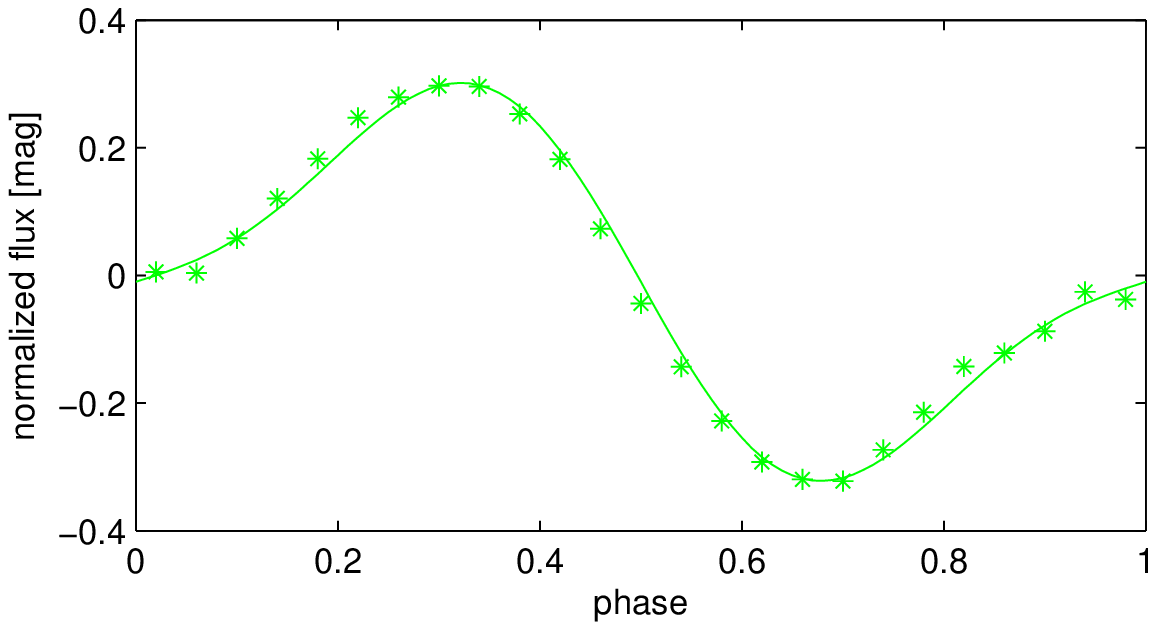}
\caption{First three principal components. The first component (top) represents a single-wave light curve, the second component (middle) a double-wave light curve, and the third component (bottom) expresses antisymmetry. The points are real principal components and the solid lines are fits using Eq. \ref{model} with $A_3=0$ for the first two components and $A_1=A_2=0$ for the third one.}\label{pca}
\end{center}
\end{figure}

As PCA requires common data points for all light curves (phase points in this case), we divided all measurements for any given star into 25 phase bins. Every phase bin of each star is then represented by its average value of intensity in the $V$ filter. Applying PCA, we acquired full sets of principal components. These components were fitted by harmonic polynomials of second order, and fit parameters and their errors were obtained. We discovered that only the first five components are meaningful, as their fitted parameters have higher values than their errors multiplied by three. The eigenvalues of these components are 1.89, 0.17, 0.16, 0.03, and 0.03. For simplicity we decided to work only with the first three principal components. These are shown in Fig. \ref{pca}.

\begin{figure}
\begin{center}
\includegraphics[width=0.4\textwidth]{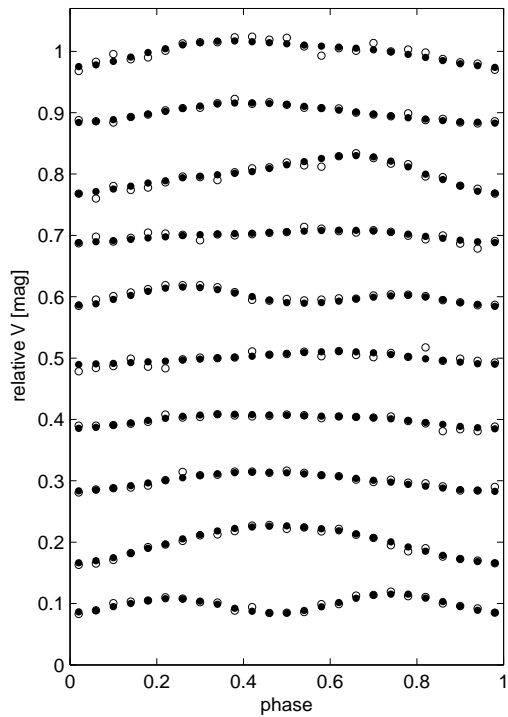}
\caption{Binned light curves of the first ten stars of our sample. The open circles represent averages of real measurements in a particular bin. The filled circles indicate the fit obtained as a linear combination of the first three principal components. The $y$-axis describes the relative changes in $V$ flux for each light curve. The light curves for different stars have been vertically shifted for clarity.}\label{pcaLC}
\end{center}
\end{figure}

 The first component represents a typical mCP-star light curve, which is a symmetric single wave with a relatively sharp maximum and very flat minimum. This light-curve profile can be simulated by assuming one bright photometric spot centered at phase zero. The second principal component is a symmetric double-wave light curve with uneven minima. The third component represents an antisymmetric light curve with the inflection points at phases 0.0 and 0.5, expressing the antisymmetry of light curves. These three components are sufficient to adequately describe all light curves, as indicated by Fig. \ref{pcaLC}, which illustrates the binned light curves of the first ten stars of our sample overlaid with the smoothed data obtained as a linear combination of the first three principal components.

\subsection{Phenomenological model}

\begin{figure}
\begin{center}
\includegraphics[width=0.45\textwidth]{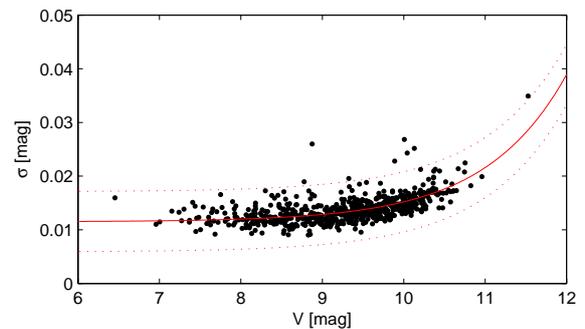}
\caption{Standard deviation of the fit ($\sigma$) for all sample stars. The full red line represents the fit. Dotted red lines define the $3\sigma_{\mathrm{exp}}$ interval.}\label{std}
\end{center}
\end{figure}

As has been pointed out, PCA shows that the model function can be satisfactorily described by only the first three principal components, with the first two components being symmetrical and the third one antisymmetrical with regard to phase zero. All of the components can be well approximated by a harmonic polynomial of the second order \citep[see also][]{2007astro.ph..3521M,2015A&A...581A.138B}. This allows us to express the ASAS-3 mCP-star light curves in a simple form with sufficient accuracy (which is quantitatively evaluated in Fig. \ref{std}), using only three free parameters describing their shape:
\begin{eqnarray}
F&=&F_0+F_1(\phi)+F_2(\phi)+F_3(\phi)=\nonumber\\
&=&F_0+A_1\cos{(2\pi\phi)}+A_2\cos{(4\pi\phi)}+\nonumber\\
&&+\,A_3\frac{2}{\sqrt{5}}\left[\sin{(2\pi\phi)}-\frac{1}{2}\sin{(4\pi\phi)}\right],\label{model}
\end{eqnarray}
where
\begin{eqnarray}
&\displaystyle \phi=\varphi-\varphi_0;\qquad\varphi=\mathrm{frac}\left(\frac{t-M_0}{P}\right).
\end{eqnarray}
Here, $F$ represents the apparent magnitude in the $V$ filter and $\varphi$ is the phase. We used the period values ($P$) and times of maximum ($M_0$) derived by \citet{2015A&A...581A.138B} and \citet{2016AJ....152..104H}. Calculations were complicated by the fact that, employing this model function, we need to work with a nonlinear regression. However, we can assume that the nonlinear parameter $\varphi_0$ has a value near zero, as the light curves have their photometric maxima centered at phase zero, defined by the parameter $M_0$.

In this model, the parameters $A_1$, $A_2$, and $A_3$ represent, respectively, the semi-amplitudes of the single-wave, the double-wave, and the antisymmetric terms of the light curve. The special form of the antisymmetric part results from two factors: (a) the first derivative in zero phase has to be zero to ensure the position of the maximum; (b) the value of parameter $A_3$ is modified by the coefficient $2/\sqrt{5}$, which normalizes it to the other two parameters. The normalization has been derived from integration of partial function squares:
\begin{eqnarray}
\int_0^1{F_1^2(\phi)\diff \phi}&=&A_1^2\int_0^1{\cos^2{(2\pi\phi)}\diff \phi} = \frac{A_1^2}{2}\\
\int_0^1{F_2^2(\phi)\diff \phi}&=&A_2^2\int_0^1{\cos^2{(4\pi\phi)}\diff \phi} = \frac{A_2^2}{2}\\
\int_0^1{F_3^2(\phi)\diff \phi}&=&A_3^2\frac{4}{5}\int_0^1\left[\sin^2{(2\pi\phi)}-\sin{(2\pi\phi)}\sin{(4\pi\phi)}+\right.\nonumber\\
&&+\left.\frac{1}{4}\sin^2{(4\pi\phi)}\right]\diff \phi = \frac{A_3^2}{2}
.\end{eqnarray}
As the parameters $A_1$, $A_2$, and $A_3$ are normalized, we can introduce an effective semi-amplitude $A$, where $A=\sqrt{A_1^2+A_2^2+A_3^2}$ ,  in agreement with a general definition of the quantity in \citet{2007AN....328...10M}.

With the prepared model functions described by Eq. \ref{model}, we were able to fit the original data in unbinned form and derive values of the parameters $F_0$, $A_1$, $A_2$, and $A_3$ for all analyzed stars, which are given in the Appendix (Section \ref{appendix_fix_parameters}).

\subsection{Peculiarities of ASAS-3 data}

\begin{figure}
\begin{center}
\includegraphics[width=0.45\textwidth]{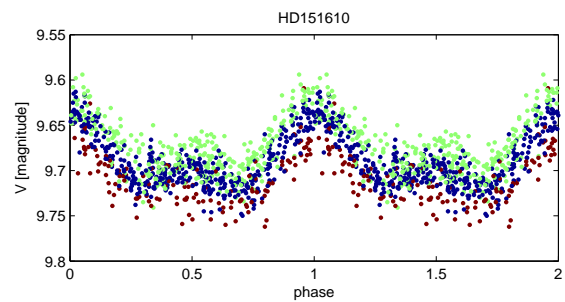}
\caption{Phased light curve of HD\,151610. This illustrates the occasional shifts in the zero level of mean brightness between different ASAS-3 data sets (indicated by the use of different colours).}\label{error2}
\end{center}
\end{figure}

ASAS-3 data suffer from occasional inhomogeneities, which cause shifts in the mean brightness level of the observations (see e.g., the case of HD\,151610 depicted in Fig. \ref{error2}). This problem can be solved if different values of the parameter $F_0$ in Eq. \ref{model} are assumed for the different data sets $(F_{0i})$. However, this procedure has to be performed after the elimination of outliers, which we identified using a 3.5\,$\sigma$ interval. In addition, data sets with less than four measurements need to be removed because they contain less points than the number of free parameters. Our analysis showed that, in some extreme and rare cases, the zero-point shifts between different data sets amount to about 0.05\,mag.

The described phenomenological model enables us to fit the ASAS-3 data with high accuracy using at least four parameters, which depends on the number of parameters $F_{0i}$. The standard deviations of the fit increase from 0.01\,mag for stars of the seventh magnitude to approximately 0.03\,mag for objects of the eleventh magnitude. These values can be interpreted as the intrinsic accuracy of ASAS-3 data and correspond well to the values derived by \citet{2014IAUS..301...31P}. The relation between the standard deviation of the data fitted by the model described in Eq. \ref{model} and the observed brightness in $V$ was fitted by the simple exponential $y=a+b\exp(x)$. Only 1\,\% of the fitted light curves exhibit standard deviations of the fit outside the $3\sigma_{\mathrm{exp}}$ interval (see Fig. \ref{std}), where $\sigma_{\mathrm{exp}}$ is the standard deviation derived from an exponential fit.

\begin{figure}[h]
\begin{center}
\includegraphics[width=0.45\textwidth]{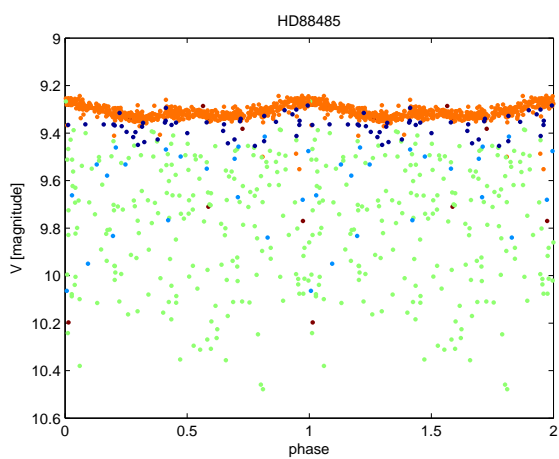}
\caption{Phased ASAS-3 light curve of HD\,88485, exhibiting obvious blending issues. Different colors represent different sets of measurements. While the orange data denote the real light curve of the star, the very different green data are likely influenced by a background object.}\label{error1}
\end{center}
\end{figure}

Photometric data for different stars from the ASAS survey are identified by a designation consisting of right ascension and declination (e.g., the light curve of HD\,10081, the very first star from our sample, is designated ASAS J013615-6815.1). This identification, however, is not always sufficiently precise, and data from different stars can be blended, especially in high-density fields. In fact, this is the case for several stars of our sample, which in consequence exhibit inappropriately high standard deviations. To identify blended data, we separated the measurements from different data sets. Results for a prototypical star are shown in Fig.\,\ref{error1}. In this particular case, the source coordinates of the different data sets are slightly different. The scatter in some of these data sets is significantly higher than the intrinsic accuracy of ASAS-3 data, which indicates blending issues and, possibly, the variability of another star in the same field of view.

\subsection{Results}

The nonlinear regression provides us with values of parameters and corresponding uncertainties. These values are derived under the assumption of a normal distribution of the fit residuals, and thus their determination does not have to be exact in cases of strongly nonlinear model functions. We therefore used the bootstrapping technique to determine corresponding error values and to verify the regression results. The results gained from both methods differ only slightly. Error values $\delta A_1$, $\delta A_2$, $\delta A_3$, and $\delta\varphi_0$ derived from the bootstrapping procedure reach, on average, 100\,, 98\,, 100\, and 93\,\% of the values derived from the regression analysis.

As the values of the standard deviations for the corrected source data (i.e., data without blending issues) follow the intrinsic accuracy of the ASAS-3 data (Fig. \ref{std}), we can assume that the employed phenomenological model fits the data with high accuracy. The key outputs are the values of the model parameters. Obtained values of zero phase $\varphi_0$ are in agreement with the expectations; only in 7\,\% of all cases do the values lie outside the interval $[-0.05,0.05]$.

Let us assume that the values of the parameters $A_1$, $A_2$, and $A_3$ in Eq. \ref{model} are significant only if their absolute values exceed three times their errors $\delta A_i$ as calculated using the bootstrapping technique. As stated above, we employ the term \textit{double wave} if a light curve fitted with Eq. \ref{model} contains two minima and has a significant value of parameter $A_2$. Similarly, the term \textit{single wave} is reserved for stars with light curves characterized by a single minimum and a significant value of parameter $A_1$. The number of minima was determined individually for each given star by mathematically inspecting the local minima of the fitted light curve. These definitions of significance and wave character are used in the following to describe the different shapes of the mCP-star light curves.

\begin{figure}
\begin{center}
\includegraphics[width=0.35\textwidth]{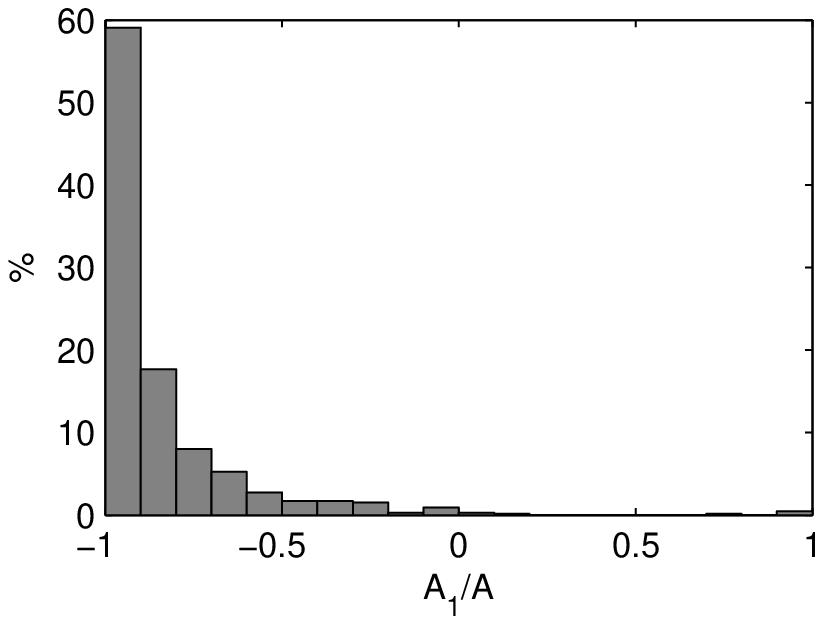}
\includegraphics[width=0.35\textwidth]{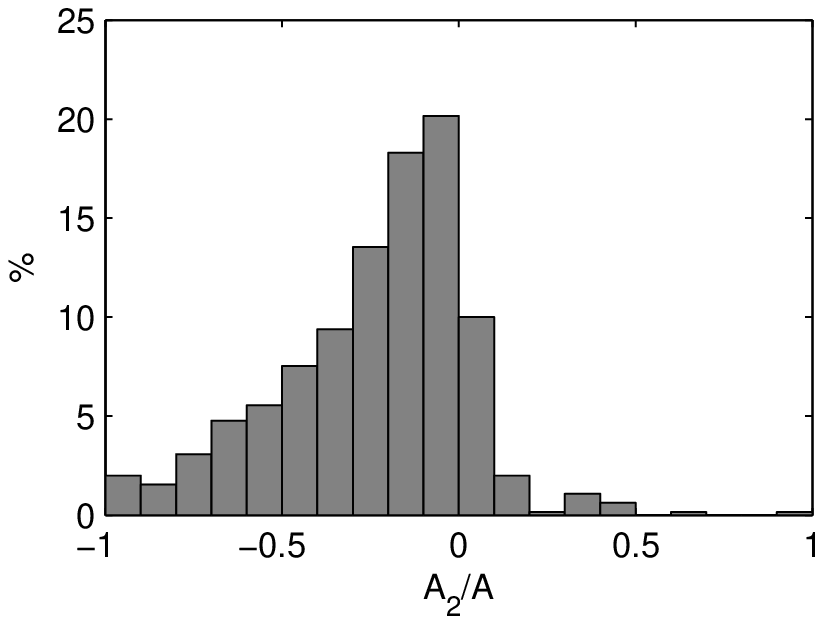}
\includegraphics[width=0.35\textwidth]{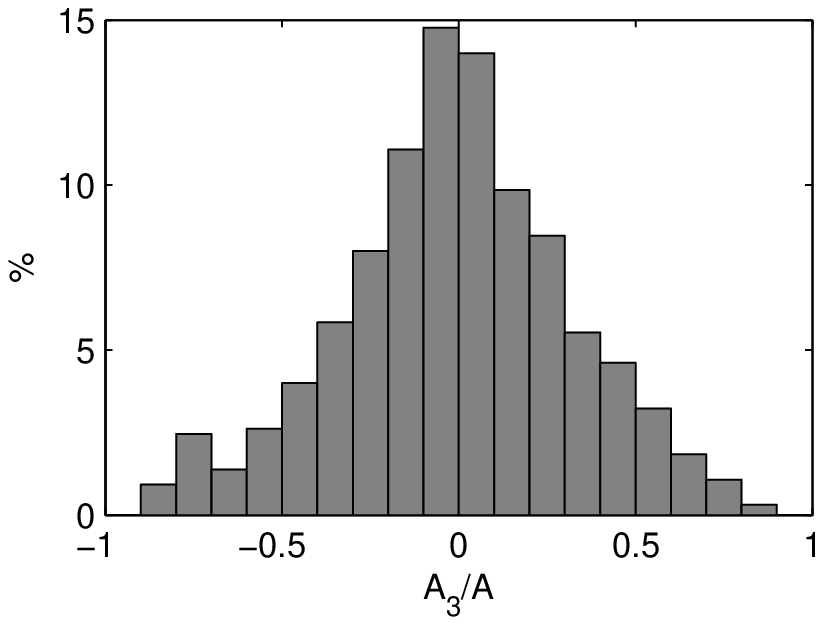}
\caption{Probability histograms of the normalized parameters $A_1$, $A_2$, and $A_3$.}\label{hist}
\end{center}
\end{figure}

\begin{figure}
\begin{center}
\includegraphics[width=0.49\textwidth]{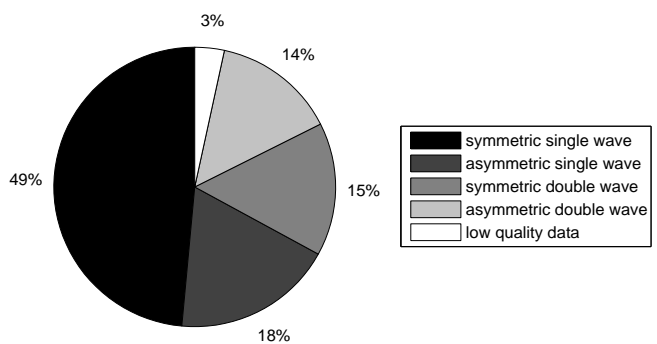}
\caption{{ Statistical distribution of the shape of light curves of our sample of 650 mCP stars.}}\label{pie}
\end{center}
\end{figure}

The most common shape for a light curve among the photometrically variable mCP stars of our sample is the single wave, with 67\,\% of our sample stars showing this characteristic. This is illustrated in the upper panel of Fig. \ref{hist} (cf. also Fig. \ref{pie}). It is furthermore apparent that the parameter $A_1$ is, in general, the main source of the effective semi-amplitude $A$. Furthermore, $A_1$ is significant in 98\,\% of all cases. The values tend to be negative as the light curve maxima are centered at $\varphi_0=0$. This might seem counterintuitive; however, we need to remember that the brightness maximum is defined by minimum magnitude. Also, 49\,\% of light curves show a symmetric single wave, while 18\,\% are characterized by an asymmetric single wave.

The middle panel of Fig. \ref{hist} shows the histogram of the normalized parameter $A_2$. Significant values of $A_2$ have been found for 46\,\% of our sample stars. Also, 30\,\% of all light curves are characterized by double waves; 15\,\% of our sample stars show symmetric double waves, while 14\,\% show asymmetric double waves; see Fig. \ref{pie} (the apparent inconsistency in expected equality of summed rate of symmetric and asymmetric double waves (29\,\%) and all double waves (30\,\%) is caused by rounding). It appears that 1\,\% of stars possess light curves with identical primary and secondary minima, in which case $A_1$ and $A_3$ are insignificant. However, we caution that the periods were derived from photometric observations only, and some period values may actually represent twice the true value, which is likely to have influenced this result. Without additional (spectroscopic) data, it is sometimes difficult to determine whether a light curve shows a double wave with identical minima or a single wave. Nevertheless, as the following analysis shows, double-wave light curves with identical minima are very rare. Therefore, the above described issue, if at all, applies only to a small portion of our sample stars and will not have influenced our results in a significant way.

The histogram of the values of the normalized parameter $A_3$, which seems to be significant in 36\,\% of all cases, appears to be symmetrically centered at zero (Fig. \ref{hist}, bottom panel), which supports the claim that the observed antisymmetry is random. However, the distribution is slightly left-skewed ($-0.11\pm0.09$).

The statistical distribution of the light curve shape according to wave character and symmetry is summarized in Fig. \ref{pie}, while types and definitions are contained in Table \ref{classification}. It becomes obvious that among the symmetric light curves, the single-wave character is highly dominant (77\,\% vs. 55\,\% among antisymmetric light curves). It must be noted that the category of light curves with two minima and insignificant $A_2$ (3\,\%) suffers from observations with high dispersion, and the corresponding values of $A_2$ are lower than their estimated errors. This is indicative of low-quality data, and we consequently did not include these light curves in the category of double-wave light curves. They are indicated as \textit{low-quality data} in Fig. \ref{pie}.

\begin{table}[h]
\begin{center}
\begin{tabular}{l|lll}
  \hline
  Type & \multicolumn{2}{c}{Conditions} & n\\
  \hline
  symmetric SW & $|A_1|>3\delta A_1$ & $|A_3|<3\delta A_3$ & 1 \\
  symmetric DW & $|A_2|>3\delta A_2$ & $|A_3|<3\delta A_3$ & 2 \\
  asymmetric SW & $|A_1|>3\delta A_1$ & $|A_3|>3\delta A_3$ & 1 \\
  asymmetric DW & $|A_2|>3\delta A_2$ & $|A_3|>3\delta A_3$ & 2 \\
  low-quality data & \multicolumn{3}{l}{all other data} \\
  \hline
\end{tabular}
\end{center}
\caption{Classification of light curves. SW = single wave, DW = double wave, n = number of local minima in the light curve.}\label{classification}
\end{table}

\section{Simulations of symmetric spot distributions}

If we assume that the main factor responsible for the distribution of chemical spots (which, as outlined in Sect. \ref{intro}, correspond to the photometric spots; cf. \citealt{2007A&A...470.1089K} and \citealt{2010A&A...524A..66S}) is the magnetic field, we can expect some kind of correlation between the distribution of the spots and some magnetic field quantity. The recent study of \citet{2015MNRAS.454.3143A} shows that large over-abundances are expected near the magnetic equator because this region boasts almost horizontal magnetic field lines, which seems to be an important factor for the creation of surface chemical spots. We here try to confirm the anticipated relation between magnetic field and photometric spots employing the following model simulations.

\begin{figure}
\begin{center}
\includegraphics[width=0.35\textwidth]{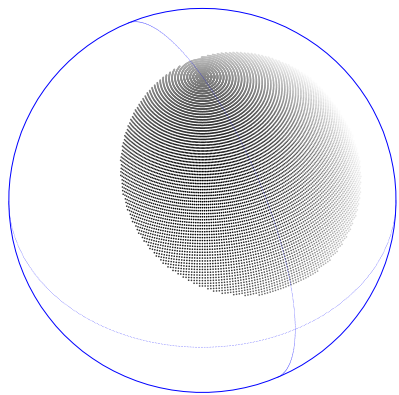}
\caption{Exemplary simulation of a model spot with parameters $\alpha=0.7$~rad, $\delta=1.0$~rad, $i=0.7$~rad.}\label{point_model}
\end{center}
\end{figure}

The basic assumption is that the magnetic field of the model star is central, axially symmetric, and dipolar, with an inclination between rotational and magnetic axis $\beta$. We can then assume that the flux distribution can be represented by two circular spots located around the magnetic poles. We tried to simulate this by using circular spots of angular radius $\alpha=0.7$ rad (spots of medium size) on a surface grid system assuming random orientation of both rotational and magnetic axes. The surface grid was defined as a system of stellar meridians and parallels with a uniform grid step value of $\pi/150$\,rad. For each grid node, the value of flux density was calculated and the total flux was integrated over the whole surface. An exemplary simulation is shown in Fig. \ref{point_model}. Parameter $\delta$ expresses the stellar latitude of the center of the spot and $i$ is the  inclination of the rotational axis towards the observer. Random orientation of rotational and magnetic axes was established by choosing $\delta=\arcsin(\mathrm{rand}(0,1))$ and $i=\arccos(\mathrm{rand}(0,1))$, where $\mathrm{rand}(0,1)$ picks random numbers between 0 and 1 with uniform distribution. The value of the employed rotational phase step is 0.025. In our calculations, we took into account the limb-darkening formula $I(\rho)=I_0[1-q(1-\cos{\rho})]$ \citep{2011A&A...530A..65N}, where $q$ is a linear limb-darkening coefficient (here we use the value $q = 0.5$), and $\rho$ is the angular distance from the center of the stellar disk.

At first, we assumed two completely dark spots centered at opposite longitudes and stellar latitude $\pm\delta$. The inclination of the stellar rotational axis $i$ is the angle between the rotational axis and the line of sight of the observer. As a result, we obtained a light curve fitted by the model function described in Eq. \ref{model}.

This simulation results in light curves that are always symmetric. Therefore, we removed the redundant antisymmetric part from Eq. \ref{model}. From light-curve fits, we obtained values of the parameters $A_1$ and $A_2$. In this way, a thousand random light curves were created.

To simplify the process, we assumed dark spots in our simulations, despite the fact that the photometric spots on mCP stars are usually bright in the optical region \citep{2015MNRAS.451.2015O}. However, this should not pose a problem as the shape of a light curve containing a bright spot is simply a vertically inverted image of a light curve with a dark spot. This, therefore, does not influence the single- or double-wave character of the simulated light curves.

\begin{figure}[h]
\begin{center}
\includegraphics[width=0.35\textwidth]{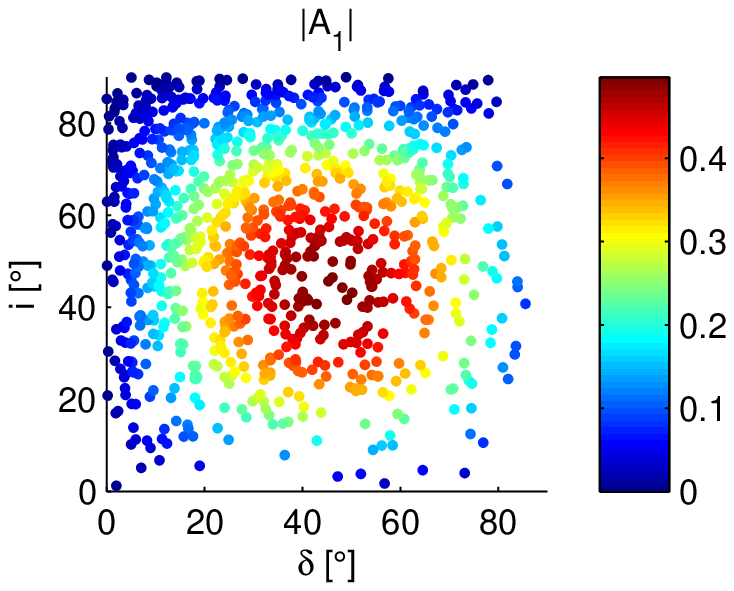}
\includegraphics[width=0.35\textwidth]{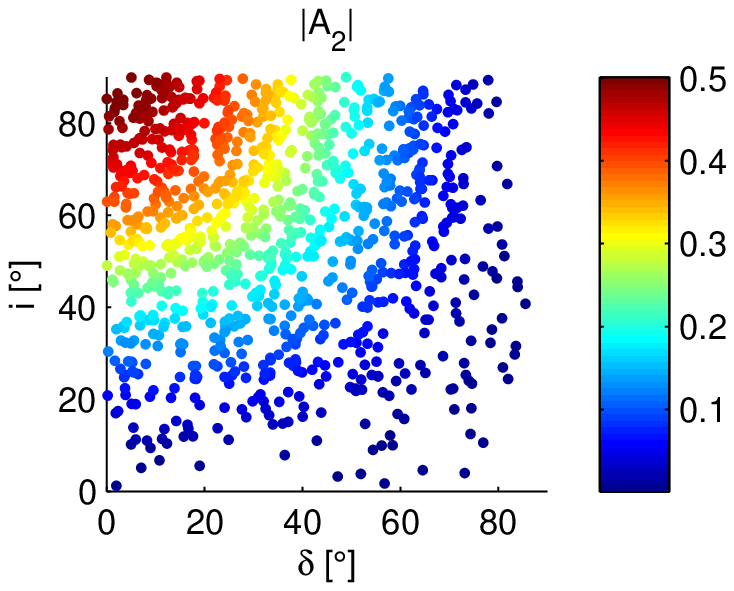}
\includegraphics[width=0.35\textwidth]{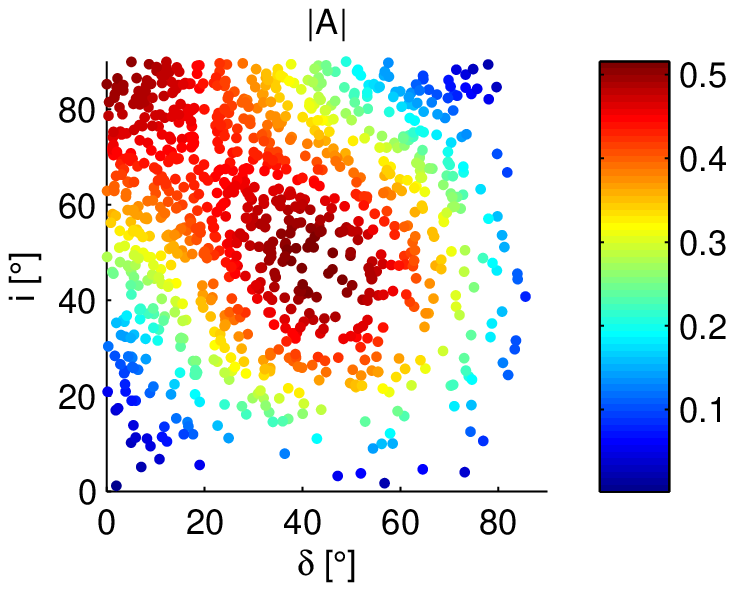}
\caption{Dependence of the parameters $A_1$, $A_2$, and $A$ on the inclination of the rotation axis $i$ and the declination of the spots $\delta$. Values are expressed in magnitudes, as indicated on the right side. The simulations have been based on the assumption of two completely dark spots with radii of 0.7 rad that are located opposite each other on the star.}\label{sim}
\end{center}
\end{figure}

Figure \ref{sim} illustrates the dependence of the parameters $A_1$, $A_2$, and $A$ on the inclination of the rotation axis and the declination of the spots. We can see that the highest values of $A_1$ are derived for the combination $[\delta,i]=[45\degr,45\degr]$ and the highest values of $A_2$ for the combination $[\delta,i]=[0\degr,90\degr]$. The main result of this simulation is that double-wave light curves should occur with a probability of 79\,\%, which does not agree with the observed 77\,\% of single-wave light curves among the ASAS-3 mCP stars with symmetric light curves. There are at least two main factors which may have influenced these results. First, a double wave with identical minima can be misclassified as a single wave with half the true period. However, the simulation predicts that less than 9\,\% of cases satisfy the relation $|A_1/A_2|<0.1$, which we assume to be a satisfying condition for the identification of light curves with equal minima. Second, the spots are not completely dark, and some fainter spots will not have been detected. If we remove half of the cases with the lowest effective semi-amplitudes $A$, the occurrence rate of double-wave light curves reaches 86\,\%.

\begin{table}[h]
\begin{center}
\begin{tabular}{l|ccc}
  \hline
  Spot type & $\alpha$ [rad] & Rate [\%] & $A_{\mathrm{max}}$ [mag]\\
  \hline
  completely dark & 0.3 & 74-83 & 0.11 \\
  completely dark & 0.7 & 79-86 & 0.52 \\
  completely dark & 1.2 & 87-95 & 1.00 \\
  partially dark & 0.7 & 74-88 & 0.62 \\
  \hline
\end{tabular}
\end{center}
\caption{Simulation results for all studied cases. The columns describe, respectively, the spot type (two dark spots vs. one dark spot and one spot of random opacity), the angular radius of the spots $\alpha$, the occurrence rate of double-wave light curves, and the maximum values for the effective semi-amplitude $A$. Values of the occurrence rate of the double-wave light curves are given with a left limit case that includes all 1000 simulation runs and a right limit case where only the 500 simulations with the highest resulting values of $A$ are taken into account.}\label{cases}
\end{table}

We ran similar simulations assuming spots with a radius of 0.3\,rad, which resulted in double-wave light curves appearing in 74\,\% of cases up to 83\,\% after removal of half of the stars with the lowest effective amplitudes. However, for such small spots, values of $A$ only reach up to 0.1\,mag, even if we assume the spots are completely dark. It is therefore unlikely that we observe spots of this size. Simulations assuming large spots of radius 1.2\,rad resulted in 87\,\% of double-wave light curves (up to 95\,\% after removal of the lowest-amplitude objects). In addition, we carried out simulations assuming two spots with a radius of 0.7\,rad, one being totally dark and the other one of random opacity. Similar results were obtained, with double wave light curves appearing in 74\,\% of cases (up to 88\,\% after removal of half of the stars with the lowest effective amplitudes). All studied cases are summarized in Table \ref{cases}.

In summary, the results of our simulations suggest that, with the given assumptions, the light curves of mCP stars tend to be double-wave rather than single-wave. However, this is not compatible with the observations, which indicates that the model assuming similar distributions of magnetic field and photometric spots is not adequate. This is in agreement with the results of several studies employing Doppler mapping of magnetic fields in mCP stars and a comparison to the chemical abundance maps of different elements \citep[e.g.,][]{2002A&A...389..420K,2014A&A...565A..83K,2015A&A...574A..79K}.

The symmetry of the simulated light curves is a result of the chosen model. Real light curves are not always symmetric and the presence of nonzero parameter $A_3$ can be explained by noncircular spots or shifted spot location as compared to the simulations, which assumed that the spots are located opposite to each other.

\section{Conclusions}

Using archival time-series data from the ASAS-3 survey, we analyzed the sample of 650 ACV variables and candidates by \citet{2015A&A...581A.138B} and \citet{2016AJ....152..104H}. We focused on the study of the general shapes of their light curves as relating to their symmetry and single- or double-wave character. The results show that 67\,\% of the light curves are single-wave, 30\,\% are double-wave, and asymmetry is found in 35\,\% of cases. Among the symmetric light curves, the percentage of single-wave light curves is even higher (77\,\%). We did not observe any kind of systematic antisymmetry.

We assumed a theoretical model of a symmetric dipole magnetic field with two opposite circular spots that follow the symmetry of the magnetic field and are hence located around the magnetic poles. Using this model, we made simulations assuming random positions of spots and random inclination of the rotational axis. Assuming completely dark, medium-sized identical spots with angular radii of 0.7\,rad, we find an occurrence rate of 79\,\% for double-wave light curves. If we ignore the simulated light curves with the lowest amplitudes to take into account the lesser probability of their detection, the occurrence rate rises to 86\,\%. In cases of smaller spots with angular radii of 0.3\,rad, similar results were obtained: 74\,\% of the light curves are double-wave, rising up to 83\,\% if low-amplitude light curves are ignored. However, the simulations show that the smaller spots frequently result in rotational modulation with amplitudes too low to be detectable in ASAS-3 data. Simulations for large spots with angular radii of 1.2 rad result in 87\,\% of light curves being  double-wave, and up to 95\,\% after removal of half of the faintest cases.

In summary, the results obtained from our simulations are in contradiction with the observed light curves of  mCP stars. While we have assumed quite simple initial conditions for our spot-modelling attempts, we do not see any reason to believe that our results are influenced by significant systematic error. In disagreement with the current understanding, we thereby conclude that the distribution of photometric spots does not follow the magnetic field topology.

\begin{acknowledgements}
This research was supported by grant GA \v{C}R 16-01116S.
\end{acknowledgements}

\bibliographystyle{aa}
\bibliography{bib}

\begin{appendix}

\section{Fit parameters and ephemerides of the sample of 650 mCP stars.} \label{appendix_fix_parameters}

\longtab[1]{
\begin{longtable}{lrrrrrrrrcr}
\caption{Fit parameters and ephemerides of the 650 mCP stars. From left to right, the columns denote: the object identification (HD number or other conventional ID); the parameters $A_1$, $A_2$, and $A_3$ with their errors $\delta A_1$, $\delta A_2$, and $\delta A_3$; the standard deviation $\sigma$ derived from Eq. \ref{model}; the period value $P$ and time of maximum light $M_0$ taken from \citet{2015A&A...581A.138B} and \citet{2016AJ....152..104H}; and the wave number $n$, which indicates (1) single-wave or (2) double-wave character of the light curve.}\\
\hline
\hline
\multicolumn{1}{c}{Star} & \multicolumn{1}{c}{$A_1$} & \multicolumn{1}{c}{$A_2$} & \multicolumn{1}{c}{$A_3$} & \multicolumn{1}{c}{$\delta A_1$} & \multicolumn{1}{c}{$\delta A_2$} & \multicolumn{1}{c}{$\delta A_3$} & \multicolumn{1}{c}{$\sigma$} & \multicolumn{1}{c}{P} & \multicolumn{1}{c}{$M_0-2450000$} & n \\
\multicolumn{1}{c}{} & \multicolumn{1}{c}{[mag]} & \multicolumn{1}{c}{[mag]} & \multicolumn{1}{c}{[mag]} & \multicolumn{1}{c}{[mag]} & \multicolumn{1}{c}{[mag]} & \multicolumn{1}{c}{[mag]} & \multicolumn{1}{c}{[mag]} & \multicolumn{1}{c}{[d]} & \multicolumn{1}{c}{[d]} & \\\hline
\endfirsthead
\caption{Continued.} \\
\hline
\multicolumn{1}{c}{Star} & \multicolumn{1}{c}{$A_1$} & \multicolumn{1}{c}{$A_2$} & \multicolumn{1}{c}{$A_3$} & \multicolumn{1}{c}{$\delta A_1$} & \multicolumn{1}{c}{$\delta A_2$} & \multicolumn{1}{c}{$\delta A_3$} & \multicolumn{1}{c}{$\sigma$} & \multicolumn{1}{c}{P} & \multicolumn{1}{c}{$M_0-2450000$} & n \\
\multicolumn{1}{c}{} & \multicolumn{1}{c}{[mag]} & \multicolumn{1}{c}{[mag]} & \multicolumn{1}{c}{[mag]} & \multicolumn{1}{c}{[mag]} & \multicolumn{1}{c}{[mag]} & \multicolumn{1}{c}{[mag]} & \multicolumn{1}{c}{[mag]} & \multicolumn{1}{c}{[d]} & \multicolumn{1}{c}{[d]} & \\\hline
\hline
\endhead
\hline
\endfoot
\hline
\endlastfoot
HD10081 &       0.0004  &       -0.0129 &       -0.0027 &       0.0006  &       0.0006  &       0.0007  &       0.0132  &       1.570   &       2190.56 &       2       \\
HD26726 &       -0.0301 &       -0.0010 &       -0.0008 &       0.0009  &       0.0009  &       0.0018  &       0.0151  &       5.382   &       1902.60 &       1       \\
HD30898 &       -0.0154 &       -0.0004 &       0.0007  &       0.0008  &       0.0008  &       0.0016  &       0.0109  &       0.782   &       2185.79 &       1       \\
HD35177 &       -0.0103 &       -0.0036 &       0.0004  &       0.0007  &       0.0007  &       0.0007  &       0.0093  &       0.528   &       2670.60 &       2       \\
HD36997 &       -0.0119 &       -0.0039 &       -0.0004 &       0.0008  &       0.0008  &       0.0010  &       0.0134  &       6.007   &       3101.50 &       2       \\
HD37189 &       -0.0030 &       -0.0105 &       0.0066  &       0.0006  &       0.0006  &       0.0006  &       0.0124  &       2.829   &       2115.90 &       2       \\
HD37713 &       -0.0092 &       -0.0026 &       -0.0044 &       0.0008  &       0.0008  &       0.0009  &       0.0126  &       1.501   &       1914.64 &       1       \\
HD38366 &       -0.0266 &       -0.0044 &       -0.0098 &       0.0011  &       0.0010  &       0.0013  &       0.0128  &       0.625   &       4143.68 &       1       \\
\end{longtable}
}
\end{appendix}

\end{document}